\documentclass[12pt]{article}
\usepackage{amsmath}
\usepackage{amsfonts}
\usepackage{amsthm}
\usepackage{graphicx}
\newtheorem{thm}{Theorem}[section]

\newtheorem{cor}[thm]{Corollary}
\newtheorem{lem}[thm]{Lemma}

\newtheorem{rem}[thm]{Remark}

\begin{document}

\begin{center}
{\Large On Some  Algebraic Properties of Semi-Discrete Hyperbolic
Type Equations}

\vskip 0.2cm

{Ismagil Habibullin}\footnote{e-mail: habibullin\_i@mail.rb.ru,
(On leave from Ufa Institute of Mathematics, Russian Academy of
Science, Chernyshevskii Str. , 112, Ufa, 450077, Russia)}

{Asl{\i} Pekcan}

{Natalya Zheltukhina}

{Department of Mathematics, Faculty of Science,
 \\Bilkent University, 06800, Ankara, Turkey \\}

\end{center}
\vskip 0.2cm
\begin{abstract}
Nonlinear semi-discrete equations of the form
$t_x(n+1)=f(t(n),t(n+1), t_x(n))$ are studied. An adequate algebraic
formulation of the Darboux integrability is discussed and the
attempt to adopt this notion to the classification of Darboux
integrable chains has been undertaken.
\end{abstract}

\section{Introduction}

Investigation of the class of hyperbolic type differential
equations of the form
\begin{equation}
u_{xy}=f(x,y,u,u_x,u_y) \label{hyp}
\end{equation}
has a very long history. Various approaches have been developed to
look for particular and general solutions of these kind equations.
In the literature one can find several definitions of integrability
of the equation. According to one given by G.~Darboux (see
\cite{Darboux}, \cite{GrundlandVassiliou}), equation (\ref{hyp}) is
called integrable if it is reduced to a pair of ordinary (generally
nonlinear) differential equations, or more exactly, if its any
solution satisfies the equations of the form
\begin{equation}
F(x,y,u,u_x,u_{xx},..., D_x^mu)=a(x),\quad G(x,y,u,u_y,u_{yy},...,
D_y^nu)=b(y),\label{ch}
\end{equation}
for appropriately chosen functional parameters $a(x)$ and $b(y)$,
where $D_x$ and $D_y$ are operators of differentiation with respect
to $x$ and $y$, $u_x=D_xu$, $u_{xx}=D_xu_x$ and so on. Functions $F$
and $G$ in (\ref{ch}) are called $y$- and $x$-integrals of the
equation (\ref{hyp}) respectively.

An effective criterion of Darboux integrability has been proposed by
G.~Darboux himself. Equation (\ref{hyp}) is integrable if and only
if the Laplace sequence of the linearized equation terminates at
both ends. A rigorous proof of this statement has been found only
recently  in \cite{SokolovZhiber}, \cite{AndersonKamran}. A complete
list of the Darboux integrable equations of the form (\ref{hyp}) is
given in \cite {Zhiber}.

 An alternative method of investigation and classification of the
Darboux integrable equations has been developed by A.~B.~Shabat in
\cite{ShabatYamilov}, based on the notion of characteristic Lie
algebra. Let us give a brief explanation of this notion. Begin with
the basic property of the integrals. Evidently each $y$-integral
satisfies the condition: $D_yF(x,y,u,u_x,u_{xx},..., D_x^mu)=0$.
Taking the derivative by applying the chain rule one defines a
vector field $T_1$ such that
$$ T_1F=\Big(\frac{\partial}{\partial
y}+ u_y\frac{\partial}{\partial u}+ f\frac{\partial}{\partial
u_x}+D_x(f)\frac{\partial}{\partial u_{xx}}+...\Big)F=0.
$$
So the vector field $T_1$ solves the equation $T_1F=0$. But in
general, the coefficients of the vector field depend on the
variable $u_y$ while the solution $F$ does not. This puts a severe
restriction on $F$, actually $F$ should satisfy one more equation:
$T_2F=0$, where $T_2=\displaystyle{\frac{\partial}{\partial
u_y}}$. Now the commutator of these two operators $T_1$ and $T_2$
will also annulate $F$. Moreover, for any $T$ from the Lie algebra
generated by $T_1$ and $T_2$ one gets $TF=0$. This Lie algebra is
called the characteristic Lie algebra of the equation (\ref{hyp})
in the direction of $y$. Characteristic Lie algebra in the
$x$-direction is defined in a similar way. Now by virtue of the
famous Jacobi theorem, equation (\ref{hyp}) is Darboux integrable
if and only if both of its characteristic algebras are of finite
dimension. In \cite{ShabatYamilov} and \cite{LeznovSmirnovShabat},
the characteristic Lie algebras for the systems of nonlinear
hyperbolic equations and their applications are studied.

The characteristic Lie algebra has proved to be an effective tool
for classifying nonlinear hyperbolic partial differential equations.
This concept can be extended to discrete versions of partial
differential equations (see \cite{Habibullin}).  Discrete models
have become rather popular in the last decade because of their
applications in physics and biology (see survey \cite{Zabrodin}).
The problem of classification of the discrete Darboux integrable
equations is very important and, to our knowledge, still open.
Emphasize that the notion of integrability has a various of
meanings. Different approaches and methods are applied for
classifying different types of integrable equations (see
\cite{IbragimovShabat} - \cite{Gurses4}).

In this paper we  study semi-discrete chains of the following form
\begin{equation}\label{dhyp}
t_{1x}=f(t,t_1,t_x)
\end{equation}
from the Darboux integrability point of view. Here the unknown
$t=t(n,x)$ is a function of two independent variables: one
discrete $n$ and one continuous $x$. It is assumed that
\begin{equation}\label{nonzero}
\frac{\partial f}{\partial t_x}\neq 0\, .
\end{equation}
 Subindex
means shift or derivative, for instance, $t_1=t(n+1,x)$ and
$t_x=\displaystyle{\frac{\partial}{\partial x}}t(n,x)$. Below we
use $D$ to denote the shift operator and $D_x$ to denote the
$x$-derivative: $Dh(n,x)=h(n+1,x)$ and
$D_xh(n,x)=\frac{\partial}{\partial x}h(n,x)$. For the iterated
shifts we use the subindex: $D^jh=h_j$.

Introduce now notions of the $n$-integral for the semi-discrete
chain (\ref{dhyp}).
 It is natural, in accordance with the continuous case, to call a
function $I=I(x,n,t,t_x,t_{xx},...)$ an $n$-integral of the chain
(\ref{dhyp}) if it is in the kernel of the difference operator:
$(D-1)I=0$. In other words, $n$-integral should still be
unchanged under the action of the shift operator $DI=I$, (see also
\cite{AdlerStartsev}). One can write it in an enlarged form
\begin{equation}\label{I}
I(x,n+1,t_1,f,f_x,f_{xx},...)=I(x,n,t,t_x,t_{xx},...).
\end{equation}
Notice that it is a functional equation, the unknown is taken at
two different "points". This circumstance causes the main
difficulty in studying discrete chains. Such kind problems appear
when one tries to apply the symmetry approach to discrete
equations (see \cite{NijhoffCapel}, \cite{GKP}). However, the
concept of the Lie algebra of characteristic vector fields can
serve as a basis for chains' investigation.

Introduce vector fields in the following way. Concentrate on the
main equation (\ref{I}). Evidently the left hand side of it
contains the variable $t_1$ while the right hand side does not.
Hence the total derivative of the function $DI$ with respect to
$t_1$ should vanish. In other words the $n$-integral is in the
kernel of the operator
$Y_1:=D^{-1}\displaystyle{\frac{\partial}{\partial t_1}}D$.
Similarly one can check that $I$ is in the kernel of the operator
$Y_2:=D^{-2}\displaystyle{\frac{\partial}{\partial t_1}}D^2$.
Really, the right hand side of the equation $D^2I=I$ which
immediately follows from (\ref{I}) does not depend on $t_1$,
therefore the derivative of the function $D^2I$ with respect to
$t_1$ vanishes. Proceeding this way one can easily prove that for
any  $j\geq 1$ the operator
\begin{equation}\label{definitionYj}
Y_j=D^{-j}\frac{\partial}{\partial t_1}D^j \end{equation} solves the
equation $Y_jI=0$.

So far we shifted the argument $n$ forward, shift it now backward
and use the main equation (\ref{I}) written as $D^{-1}I=I$.
Rewrite the original equation (\ref{dhyp}) in the form
\begin{equation}\label{dhypg}
t_{-1x}=g(t,t_{-1},t_x).
\end{equation}
This can be done because of the condition (\ref{nonzero}) assumed
above. In the enlarged form the equation $D^{-1}I=I$ looks like
\begin{equation}\label{Ig}
I(x,n-1,t_{-1},g,g_x,g_{xx},...)=I(x,n,t,t_x,t_{xx},...).
\end{equation}
The right side of equation (\ref{Ig}) does not depend on $t_{-1}$ so
the total derivative of $D^{-1}I$ with respect to $t_{-1}$ is zero,
i.e. the operator $Y_{-1}:=D\frac{\partial }{\partial t_{-1}}
D^{-1}$ solves the equation $Y_{-1}I=0$. Moreover, the operators
\begin{equation}\label{definitionY-j}
Y_{-j}=D^{j}\frac{\partial}{\partial t_{-1}}D^{-j}\, ,
\end{equation} $j\geq 1$, also satisfy similar conditions $Y_{-j}I=0$.

Summarizing the reasonings above one can conclude  that the
$n$-integral is annulated by any operator from the Lie algebra
$\tilde{L}_n$ generated by the operators
\begin{equation}\label{gen}
...,Y_{-2},Y_{-1},Y_{-0},Y_{0},Y_{1},Y_{2},...,
\end{equation}
where
\begin{equation}
\label{Y0Y-0} Y_0=\frac{\partial }{\partial t_{1}}\qquad
{\mbox{and}}\qquad Y_{-0}=\frac{\partial }{\partial t_{-1}}
\end{equation}
The algebra $\tilde{L}_n$ consists of the operators from the
sequence (\ref{gen}), all possible commutators, and linear
combinations with coefficients depending on the variables $n$ and
$x$. Evidently equation (\ref{dhyp}) admits a nontrivial
$n$-integral only if the dimension of the algebra $\tilde{L}_n$ is
finite. But it is not clear that the finite dimension of
$\tilde{L}_n$ is enough for existence of $n$-integrals. By this
reason we introduce another Lie algebra called the characteristic
Lie algebra $L_n$ of the equation (\ref{dhyp}) in the direction of
$n$. First we define in addition to the operators $Y_1, Y_2,...$
differential operators
\begin{equation}\label{definitionXj}
X_j=\frac{\partial}{\partial_{t_{-j}}}\end{equation} for $j\geq 1$.\\
The following theorem  (see \cite{HabibullinPekcan}) allows us to
define this characteristic Lie algebra.
\begin{thm}\label{thm1}
Equation (\ref{dhyp}) admits a nontrivial $n$-integral if and only
if the following two conditions hold:\\
1)  Linear space spanned by  the operators $\{Y_j \}_1^{\infty}$ is
of finite dimension, denote this dimension by $N$;\\
2)  Lie algebra $L_n$ generated by the operators
${Y_1,Y_2,...,Y_N,X_1,X_2,...,X_N}$ is of finite dimension. We call
$L_n$ the characteristic Lie algebra of (\ref{dhyp}) in the
direction of $n$.
\end{thm}
\begin{rem}\label{remark}
If dimension of the linear space $L_Y$ generated by $\{Y_j
\}_1^{\infty}$ is $N$ then the set $\{Y_j \}_1^{N}$ constitutes a
basis in $L_Y$.
\end{rem}

The $x$-integral and the characteristic algebra in the $x$-direction
of equation (\ref{dhyp}) are defined similar to the continuous case.
We call a function $F=F(x,n,t,t_{\pm 1},t_{\pm 2},...)$ depending on
a finite number of shifts an $x$-integral of the chain (\ref{dhyp}),
if the following condition is valid $D_xF=0$, i.e. $K_0F=0$, where
\begin{equation}\label{gc1} K_0= \frac{\partial }{\partial x}+t_x\frac{\partial
}{\partial t} +f\frac{\partial }{\partial t_1 }+g\frac{\partial
}{\partial t_{-1}} +f_1\frac{\partial }{\partial
t_2}+g_{-1}\frac{\partial }{\partial t_{-2} }+\ldots
\end{equation}
Vector fields $K_0$ and
\begin{equation}\label{gc1'}
X=\frac{\partial }{\partial t_x}
\end{equation}
as well as any vector field from the Lie algebra generated by $K_0$
and $X$ annulate $F$. This algebra is called the characteristic Lie
algebra $L_x$ of the chain (\ref{dhyp}) in the $x$-direction. The
following result is essential,  its proof can be found in
\cite{ShabatYamilov}.
\begin{thm}\label{thm1x} Equation
  (\ref{dhyp}) admits a nontrivial $x$-integral if and only if its Lie
algebra $L_x$ is of finite dimension.
\end{thm}

The article is organized as follows. In Section 2 we study the
algebra $L_n$ introduced in Theorem \ref{thm1}. Section 3 is devoted
to properties of the  Lie algebra $L_x$.
These algebras $L_n$ and $L_x$ can be used as a new classifying tool
for equations on a lattice. From this viewpoint the system of
equations (\ref{gc12}) is of special importance. Actually, the
consistency condition of this overdetermined system of "ordinary"
difference equations provides necessary conditions of the Darboux
integrability of the original equation (\ref{dhyp}). As an
illustration of efficiency of our approach in the last Section 4 we
study in details equation (\ref{dhyp}) admitting characteristic Lie
algebras $L_n$ and $L_x$ of minimal possible dimensions equal 2 and
3 respectively. It is proved that in this case the equation
(\ref{dhyp}) can be reduced to $t_{1x}=t_x+t_1-t$.

\section{Characteristic Lie Algebra $L_n$}

\noindent   The proof of the first two lemmas can be found in
\cite{HabibullinPekcan}. Here, we still present their short proofs
for the reader's convenience.
\begin{lem}
If for some integer $N$ the operator $Y_{N+1}$ is a linear
combination of the operators with less indices:
\begin{equation}\label{YM}
Y_{N+1}=\alpha_1Y_1+\alpha_2 Y_2+...+\alpha_NY_N
\end{equation}
then for any integer $j>N$, we have a similar expression
\begin{equation}
Y_{j}=\beta_1Y_1+\beta_2 Y_2+...+\beta_NY_N.
\end{equation}
\end{lem}
\noindent \textbf{Proof.} Due to the property
$Y_{k+1}=D^{-1}Y_kD$, we have from (\ref{YM})
$$
Y_{N+2}=D^{-1}(\alpha_1)Y_2+D^{-1}(\alpha_2)Y_3+...+D^{-1}(\alpha_N)(\alpha_1Y_1+...+\alpha_NY_N).
$$
Now by using induction one can easily complete the proof of the
Lemma. $\Box$
\begin{lem}\label{lemma22} The following commutativity relations take place:
$$ [Y_{0},Y_{-0}]=0,\quad [Y_{0},Y_{1}]=0, \quad [Y_{-0},Y_{-1}]=0.\quad $$
\end{lem}

\noindent {\bf Proof.} The first of the relations is evident (see
the definition (\ref{Y0Y-0}) of $Y_0$ and $Y_{-0}$). In order to
prove two others find the coordinate representation of the operators
$Y_{1}$ and $Y_{-1}$, defined by (\ref{definitionYj}) and
(\ref{definitionY-j}), acting on the class of locally smooth
functions of the variables $x,n,t,t_x,t_{xx},...$\,. By direct
computations
\begin{eqnarray*}
Y_1 H&=&D^{-1} \displaystyle \frac{d}{dt_1} D H\\
 &=&D^{-1}
\displaystyle \frac{d}{dt_1} H(t_1,f,f_x,...)\\
&=&\Big\{\frac{\partial}{\partial t}+D^{-1}\Big(\frac{\partial
f}{\partial t_1}\Big)\frac{\partial}{\partial
t_x}+D^{-1}\Big(\frac{\partial f_x}{\partial
t_1}\Big)\frac{\partial}{\partial
t_{xx}}+...\Big\}H(t,t_x,t_{xx},...)
\end{eqnarray*}
 one gets
\begin{eqnarray}\label{Y1}
Y_1=\frac{\partial}{\partial t}+D^{-1}\Big(\frac{\partial
f}{\partial t_1}\Big)\frac{\partial}{\partial
t_x}+D^{-1}\Big(\frac{\partial f_x}{\partial
t_1}\Big)\frac{\partial}{\partial t_{xx}}+D^{-1}\Big(\frac{\partial
f_{xx}}{\partial t_1}\Big)\frac{\partial}{\partial t_{xxx}}+...\,.
\end{eqnarray}
Now notice that all of the functions  $f$, $f_x$, $f_{xx}, ...$
depend on the variables $t_1$, $t$, $t_x$, $t_{xx}$, $...$ and do
not depend on $t_2$ hence the coefficients of the vector field $Y_1$
do not depend on $t_1$ and therefore the operators $Y_1$ and $Y_0$
commute. In a similar way by using the explicit coordinate
representation
$$
Y_{-1}=\frac{\partial}{\partial t}+D\Big(\frac{\partial g}{\partial
t_{-1}}\Big)\frac{\partial}{\partial t_x}+D\Big(\frac{\partial
g_x}{\partial t_{-1}}\Big)\frac{\partial}{\partial
t_{xx}}+D\Big(\frac{\partial g_{xx}}{\partial
t_{-1}}\Big)\frac{\partial}{\partial t_{xxx}}+...\, ,
$$
where $ g$ is defined by (\ref{dhypg}), one can prove that
$[Y_{-0},Y_{-1}]=0.$ $\Box$

The following statement turned out to be very useful for studying
the characteristic Lie algebra $L_n$.
\begin{lem}\label{lemma23} (1) Suppose that the vector field
$$
Y=\alpha(0)\frac{\partial }{\partial
t}+\alpha(1)\frac{\partial}{\partial {t_x}}+\alpha(2)\frac{\partial
}{\partial {t_{xx}}}+...,
$$
where $\alpha_x(0)=0$, solves the equation $[D_x,Y]=0$, then
$Y=\displaystyle{\alpha(0)\frac{\partial}{\partial t}}$.\\
(2)  Suppose that the vector field
$$
Y=\alpha(1)\frac{\partial}{\partial {t_x}}+\alpha(2)\frac{\partial
}{\partial {t_{xx}}}+ \alpha(3)\frac{\partial }{\partial
{t_{xxx}}}+...
$$
 solves the equation $[D_x,Y]=h Y$, where $h$ is a function of variables
 $t$, $t_x$, $t_{xx}$, $\ldots$, $t_{\pm 1}$, $t_{\pm 2}$, $\ldots$, then
$Y=0$.
\end{lem}

\noindent The proof of  Lemma \ref{lemma23}  can be easily derived
from the following formula
\begin{eqnarray}\label{DxY}
[D_x,Y]=&-(\alpha(0)f_t+\alpha(1)f_{t_x})\displaystyle{\frac{\partial}{\partial
t_1}}+(\alpha_x(0)-\alpha(1))\displaystyle{\frac{\partial}{\partial
t}}\nonumber\\&+(\alpha_x(1)-\alpha(2))\displaystyle{\frac{\partial}{\partial
{t_x}}}+
(\alpha_x(2)-\alpha(3))\displaystyle{\frac{\partial}{\partial
t_{xx}}}+...\,.
\end{eqnarray}
In the formula (\ref{Y1}) we have already given an enlarged
coordinate form of the operator $Y_1$. One can check that the
operator $Y_2$ is a vector field of the form
\begin{equation}\label{Y2}
Y_2=D^{-1}(Y_1(f))\frac{\partial}{\partial
{t_x}}+D^{-1}(Y_1(f_x))\frac{\partial}{\partial t_{xx}}
+D^{-1}(Y_1(f_{xx}))\frac{\partial}{\partial t_{xxx}}+...\,.
\end{equation}
It immediately follows from the equation $Y_2=D^{-1}Y_1D$ and the
coordinate representation (\ref{Y1}). By induction one can prove
similar formulas for arbitrary $Y_{j+1}$, $j\geq 1$:
\begin{equation}\label{Yj}
Y_{j+1}=D^{-1}(Y_j(f))\frac{\partial}{\partial
t_x}+D^{-1}(Y_j(f_x))\frac{\partial}{\partial t_{xx}}
+D^{-1}(Y_j(f_{xx}))\frac{\partial}{\partial t_{xxx}}+...\,.
\end{equation}
\begin{lem}\label{lemma24}  For any $n\geq 0$, we have
\begin{equation}\label{maincommrel}
[D_x,Y_n]=-\displaystyle \sum_{j=0}^nD^{-j}(Y_{n-j}(f))Y_j\, .
\end{equation}
In particular,
\begin{equation}\label{DxY0}
[D_x, Y_0]=-Y_0(f)Y_0\, ,
\end{equation}
\begin{equation}
[D_x, Y_1]=-Y_1(f)Y_0-D^{-1}(Y_0(f))Y_1\, .
\end{equation}
\end{lem}
\noindent \textbf{Proof.} We have,
\begin{eqnarray*}
[D_x,Y_0]H(t,t_1,t_x,t_{xx},...)&=&D_xH_{t_1}-Y_0D_xH\\
&=&(H_{tt_1}t_x+H_{t_1t_1}{t_1}_x+...)-\frac{\partial}{\partial_{t_1}}(H_tt_x+H_{t_1}{t_1}_x+...)\\
&=&-H_{t_1}f_{t_1}\\
&=&-Y_0(f)Y_0H,
\end{eqnarray*}
i.e. equation (\ref{DxY0}) holds.\\
We prove equations (\ref{maincommrel}), $n\geq 1$, by induction. By
(\ref{Y1}), (\ref{DxY}) and (\ref{DxY0}),
\begin{eqnarray*}
&&[D_x,Y_1]=-Y_1(f)\frac{\partial}{\partial
t_1}-D^{-1}(Y_0(f))\frac{\partial }{\partial t}+D^{-1}[D_x,
Y_0]f\frac{\partial }{\partial t_x} +D^{-1}[D_x,
Y_0]f_x\frac{\partial }{\partial t_{xx}}+\ldots\\
&&= - Y_1(f) Y_0-D^{-1}(Y_0(f))\frac{\partial }{\partial t}
-D^{-1}(Y_0(f)Y_0(f))\frac{\partial }{\partial t_x}-
D^{-1}(Y_0(f)Y_0(f_x))\frac{\partial }{\partial
t_{xx}}-\ldots\\
&&=-Y_1(f)Y_0-D^{-1}(Y_0(f))Y_1,
\end{eqnarray*}
 that shows that the
base of Mathematical Induction holds for $n=1$. By (\ref{Yj}),
(\ref{DxY}) and the inductive assumption, for $n\geq 1$ we have,
\begin{eqnarray*}
&&[D_x, Y_{n+1}]=-D^{-1}(Y_n(f))f_{t_x}\frac{\partial }{\partial
t_1}-D^{-1}(Y_n(f))\frac{\partial }{\partial t}+D^{-1}[D_x,
Y_n]f\frac{\partial }{\partial t_x}\\
&&+D^{-1}[D_x,
Y_n]f_x\frac{\partial }{\partial t_{xx}}+\ldots\\
&&= -Y_{n+1}(f) Y_0-D^{-1}(Y_n(f))\frac{\partial }{\partial
t}-D^{-1}\{Y_n(f)Y_0(f)+D^{-1}(Y_{n-1}(f))Y_1(f)\\
&&+ \ldots+D^{-n}(Y_0(f))Y_n(f)\}\frac{\partial }{\partial
t_x}+\ldots\\
&&=-Y_{n+1}(f)Y_0-D^{-1}(Y_{n}(f))Y_1-D^{-2}(Y_{n-1}(f))Y_2-\ldots-D^{-n-1}(Y_0(f))Y_{n+1}
,
\end{eqnarray*}
that finishes the proof of the Lemma. $\Box$

\begin{lem} Lie algebra  generated by the
operators $Y_1,Y_2,Y_3,...$ is commutative.
\end{lem}
\noindent \textbf{Proof.}  By Lemma \ref{lemma22}, $[Y_1,Y_0]=0$.
The reason for this equality is  that the coefficients of the vector
field $Y_1$ do not depend on the variable $t_1$. They might depend
only on $t_{-1}$, $t$, $t_x$, $t_{xx}$, $t_{xxx}$, $\ldots$.
 The coefficients of the vector field $Y_2$ being of the
form $D^{-1}(Y_1(D_x^jf))$ (see (\ref{Y2})) also do not depend on
the variable $t_1$. They might depend only on $t_{-2}$, $t_{-1}$,
$t$, $t_x$, $t_{xx}$, $t_{xxx}$, $\ldots$. Therefore, we have $
[Y_2,Y_0]=0. $
 Continuing this reasoning we see that for any $n\geq 1$
the commutativity relation
$$
[Y_n,Y_0]=0
$$
takes place.
 \noindent Consider now the commutator $[Y_n,Y_{n+m}]$,
$n\geq 1$, $m\geq 1$. We have,
\begin{eqnarray*}
[Y_n,Y_{n+m}]=[D^{-n}Y_0D^n,D^{-(n+m)}Y_0D^{n+m}]
=D^{-n}[Y_0,Y_m]D^n=0,
\end{eqnarray*}
that finishes the proof of the Lemma. $\Box$

\begin{lem}\label{lemma26} If the operator $Y_2=0$ then $[X_1,Y_1]=0$.
\end{lem}
\noindent \textbf{Proof}. By (\ref{Y2}), $Y_2=0$ implies that
$Y_1(f)=0$. Due to (\ref{Y1}), $Y_1(f)=0$ means that
$f_t+D^{-1}(f_{t_1})f_{t_x}=0$ and, therefore, $D^{-1}(f_{t_1})$
does not depend on $t_{-1}$. Together with Lemma~\ref{lemma24} and
the fact that $[D_x, X_1]=0$  (see the definition
(\ref{definitionXj})) it allows us to conclude that
\begin{eqnarray*}
[D_x,[X_1,Y_1]]&=&-[X_1,[Y_1,D_x]]-[Y_1,[D_x,X_1]]\\
&=&-[X_1, D^{-1}(f_{t_1})Y_1]\\
&=&-D^{-1}(f_{t_1})[X_1, Y_1],
\end{eqnarray*}
i.e. $ [D_x,[X_1,Y_1]]=-D^{-1}(f_{t_1})[X_1, Y_1]$. By Lemma
\ref{lemma24}, part (2), it follows that $[X_1, Y_1]=0$.
 $\Box$
\begin{lem}\label{lemma27}
The operator $Y_2=0$ if and only if we have
\begin{equation}\label{Y2=0relation}
f_t+D^{-1}(f_{t_1})f_{t_x}=0.
\end{equation}
\end{lem}
\noindent \textbf{Proof}.  Assume $Y_2=0$.  By (\ref{Y2}),
$Y_1(f)=0$. Due to (\ref{Y1}) equality $Y_1(f)=0$ is another way of
writing
(\ref{Y2=0relation}).\\
 Conversely, assume (\ref{Y2=0relation}) holds, i.e. $Y_1(f)=0$. It follows from
(\ref{Y2}) that $Y_2(f)=0$. Due to Lemma \ref{lemma24}, we have
$$[D_x,Y_2]=-D^{-2}(Y_0(f))Y_2$$ that implies, by Lemma
\ref{lemma23}, part (2), that $Y_2=0$. $\Box$

\begin{cor}\label{cornint}
The dimension of the Lie algebra $L_n$ associated with $n$-integral
is equal to $2$ if and only if (\ref{Y2=0relation}) holds, or the
same $Y_2=0$.
\end{cor}
\noindent \textbf{Proof}. By Theorem \ref{thm1}, the dimension of
$L_n$ is 2 if and only if $Y_2=\lambda_1X_1+\mu_1Y_1$ and
$[X_1,Y_1]=\lambda_2X_1+\mu_2Y_1$ for some $\lambda_i, \mu_i$, $
i=1, 2$.

Assume the dimension of $L_n$ is 2. Then
$Y_2=\lambda_1X_1+\mu_1Y_1$.  Since among $X_1$, $Y_1$, $Y_2$
differentiation by $t_{-1}$ is used only in $X_1$, differentiation
by $t$ is used only in $Y_1$, then $\lambda_1=\mu_1=0$. Therefore,
$Y_2=0$, or the same, by Lemma \ref{lemma27}, (\ref{Y2=0relation})
holds. \\
Conversely, assume (\ref{Y2=0relation}) holds, that is $Y_2=0$. By
Lemma \ref{lemma26}, $[X_1, Y_1]=0$. Since $Y_2$ and $[X_1, Y_1]$
are trivial linear combinations of $X_1$ and $Y_1$ then the
dimension of $L_n$ is 2. $\Box$

\section{Characteristic Lie Algebra $L_x$}

Denote by
\begin{equation}\label{gc2} K_1=[X,K_0], \quad K_2=[X, K_1], \quad \ldots,
\quad K_{n+1}=[X,K_n],\quad n\geq 1\, ,\end{equation} where $X$ and
$K_0$ are defined by  (\ref{gc1'}) and (\ref{gc1}).

\noindent It is easy to see that
\begin{equation} \label{gc3}K_1= \frac{\partial }{\partial t}
+X(f)\frac{\partial }{\partial t_1 }+X(g)\frac{\partial }{\partial
t_{-1}} +X(f_1)\frac{\partial }{\partial
t_2}+X(g_{-1})\frac{\partial }{\partial t_{-2} }+\ldots\,
,\end{equation} \begin{equation}
\label{gc4}K_n=\sum\limits_{j=1}^\infty
\left\{X^n(f_{j-1})\frac{\partial }{\partial
t_j}+X^n(g_{-j+1})\frac{\partial }{\partial t_{-j}}\right\}, \quad
n\geq 2,
\end{equation}
where $f_0:=f$ and $g_0:=g$.
\begin{lem}\label{gcl1}
 We have,
\begin{equation} \label{gc5}DXD^{-1}=\frac{1}{f_{t_x}}X, \qquad
DK_0D^{-1}=K_0-\frac{t_xf_t+ff_{t_1}}{f_{t_x}}X,\end{equation}

\begin{equation}\label{gc6} DK_1D^{-1}=\frac{1}{f_{t_x}}K_1
-\frac{f_t+f_{t_x}f_{t_1}}{f^2_{t_x}}X, \qquad DK_2D^{-1} =
\frac{1}{f^2_{t_x}}K_2-\frac{f_{t_xt_x}}{f^3_{t_x}}K_1+\frac{f_{t_xt_x}f_t}{f^4_{t_x}}X,\end{equation}

\begin{equation}\label{gc7}DK_3D^{-1}=\frac{1}{f^3_{t_x}}K_3-3\frac{f_{t_xt_x}}{f^4_{t_x}}K_2+
\left(3\frac{f^2_{t_xt_x}}{f^5_{t_x}}-\frac{f_{t_xt_xt_x}}{f^4_{t_x}}\right)K_1-\frac{f_t}{f_{t_x}}
\left(3\frac{f^2_{t_xt_x}}{f^5_{t_x}}-\frac{f_{t_xt_xt_x}}{f^4_{t_x}}\right)X.\end{equation}

\end{lem}

\noindent {\bf{Proof}}.  In the proof of this lemma whenever we
write $H$ and $H^*$ we mean functions $H(x, t, t_x, t_1, t_{-1},
t_2, t_{-2}, \ldots)$ and $H^*=D^{-1}H=H(x, t_{-1}, g, t, t_{-2},
t_1,
t_{-3}, \ldots)$. \\
  Since
$$ DXD^{-1}H=
D H^*_{t_x}= D\left\{g_{t_x}\frac{\partial H^* }{\partial g}
\right\}=D(g_{t_x})\frac{\partial H}{\partial t_x}=D(g_{t_x})XH$$
and $\displaystyle{ D(g_{t_x})}=\frac{1}{f_{t_x}}$ then $
DXD^{-1}H=\displaystyle{\frac{1}{f_{t_x}}}XH$.\\
Since
\begin{eqnarray*}
&&DK_0D^{-1}H=D\left(\frac{\partial }{\partial
x}+t_x\frac{\partial }{\partial t} +f\frac{\partial }{\partial t_1
}+g\frac{\partial }{\partial t_{-1}} +f_1\frac{\partial }{\partial
t_2}+g_{-1}\frac{\partial }{\partial t_{-2} }+\ldots
\right)H^*\\&&= D\left( \frac{\partial H^* }{\partial x}+
t_xg_{t}\frac{\partial H^*}{\partial g}+t_x\frac{\partial
H^*}{\partial t} +f\frac{\partial H^*}{\partial t_{1}
}+g\frac{\partial H^* }{\partial
t_{-1}}+gg_{t_{-1}}\frac{\partial{H^*}}{\partial g}
+\ldots\right)\\&&= \left(\frac{\partial H }{\partial x}+
t_x\frac{\partial H }{\partial t }+f\frac{\partial H}{\partial
t_1} +f_1\frac{\partial H}{\partial t_{2} }+\ldots\right )+
(t_xD(g_{t_{-1}})+f D(g_t))\frac{\partial H }{\partial t_x}
\end{eqnarray*}
and $D(g_{t_{-1}})=-\displaystyle{\frac{f_t}{f_{t_x}}}$,
$D(g_{t})=-\displaystyle{\frac{f_{t_1}}{f_{t_x}}}$ then $
DK_0D^{-1}H=K_0H-\displaystyle{\frac{t_xf_t+ff_{t_1}}{f_{t_x}}}XH$.\\
Using  formulas (\ref{gc5}) for $DXD^{-1}$,  $DK_0D^{-1}$ and the
definition (\ref{gc2}) of $K_1$  we have
\begin{eqnarray*}
&&DK_1D^{-1}=[DXD^{-1}, DK_0D^{-1}]=\left[\frac{1}{f_{t_x}}X,
K_0-\frac{t_xf_t+ff_{t_1}}{f_{t_x}}X\right]\\
&&= \frac{1}{f_{t_x}}K_1-
K_0\left(\frac{1}{f_{t_x}}\right)X-\frac{1}{f_{t_x}}X\left(\frac{t_xf_t+ff_{t_1}}{f_{t_x}}
\right)X+\frac{t_xf_t+ff_{t_1}}{f_{t_x}}\left(-\frac{f_{t_xt_x}}{f^2_{t_x}}\right)X\\
&&=\frac{1}{f_{t_x}}K_1-\frac{f_t+f_{t_x}f_{t_1}}{f^2_{t_x}}X.
\end{eqnarray*}
Using  formulas (\ref{gc5}) and (\ref{gc6}) for $DXD^{-1}$,
$DK_1D^{-1}$ and the definition (\ref{gc2}) of $K_2$  we have
\begin{eqnarray*}
&&DK_2D^{-1}=[DXD^{-1}, DK_1D^{-1}]=\left[\frac{1}{f_{t_x}}X,
\frac{1}{f_{t_x}}K_1-\frac{f_t+f_{t_x}f_{t_1}}{f^2_{t_x}}X\right]
\\&&=-\frac{f_{t_xt_x}}{f^3_{t_x}}K_1-\frac{1}{f_{t_x}}K_1\left(\frac{1}{f_{t_x}}\right)X
+\frac{1}{f^2_{t_x}}K_2-\frac{1}{f_{t_x}}X\left(\frac{f_t+f_{t_x}f_{t_1}}{f^2_{t_x}}\right)X\\
&&-
\frac{f_t+f_{t_x}f_{t_1}}{f^2_{t_x}}\frac{f_{t_xt_x}}{f^2_{t_x}}X\\
&&=\frac{1}{f^2_{t_x}}K_2-
\frac{f_{t_xt_x}}{f^3_{t_x}}K_1+\frac{f_{t_xt_x}f_t}{f^4_{t_x}}X.
\end{eqnarray*}
Using formulas (\ref{gc5}) and (\ref{gc6}) for $DXD^{-1}$,
$DK_2D^{-1}$ and the definition (\ref{gc2}) of $K_3$ we have
\begin{eqnarray*}
&&DK_3D^{-1}=[DXD^{-1},
DK_2D^{-1}]=\Big[\frac{1}{f_{t_x}}X,\frac{1}{f_{t_x}^2}K_2-
\frac{f_{t_xt_x}}{f_{t_x}^3}K_1+\frac{f_{t_xt_x}f_t}{f_{t_x}^4}X
\Big]\\
&&=
\frac{1}{f_{t_x}^3}K_3-\frac{f_{t_xt_x}}{f_{t_x}^4}K_2-\frac{2f_{t_xt_x}}{f_{t_x}^4}K_2
-\frac{1}{f_{t_x}}X\Big(\frac{f_{t_xt_x}}{f_{t_x}^3}\Big)K_1+
X\Big\{\frac{1}{f_{t_x}}X\Big(\frac{f_{t_xt_x}f_t}{f_{t_x}^4}\Big)\\
&&- \frac{1}{f^2_{t_x}}K_2\Big(\frac{1}{f_{t_x}}\Big)+
\frac{f_{t_xt_x}}{f_{t_x}^3}K_1\Big(\frac{1}{f_{t_x}}\Big)
-\frac{f_{t_xt_x}f_t}{f_{t_x}^4}X\Big(
\frac{1}{f_{t_x}}\Big)\Big\}\\
&&=\frac{1}{f_{t_x}^3}K_3-\frac{3f_{t_xt_x}}{f_{t_x}^4}K_2-
\frac{f_{t_xt_xt_x}f_{t_x}-3f_{t_xt_x}^2}{f_{t_x}^5}K_1+
\frac{f_t}{f_{t_x}}\frac{f_{t_xt_xt_x}f_{t_x}-3f_{t_xt_x}^2}{f_{t_x}^5}X\,
,
\end{eqnarray*}
that finishes the proof of the Lemma. $\Box$

\begin{lem} \label{gcl2} For any $n\geq 1$ we have,
\begin{equation}
DK_nD^{-1}=a^{(n)}_nK_n+a^{(n)}_{n-1}K_{n-1}+a^{(n)}_{n-2}K_{n-2}+\ldots+a^{(n)}_1K_1+b^{(n)}X,
\label{gc8}\end{equation} where coefficients $b^{(n)}$ and
$a^{(n)}_k$ are functions that depend only on variables $t$, $t_1$
and $t_x$ for all $k$, $1\leq k\leq n$. Moreover,
\begin{equation}\label{gc9}\left\{\begin{array}{lll}
   a^{(n)}_n&=&\displaystyle{\frac{1}{f^n_{t_x}}}, \quad n\geq 1, \\
a^{(n)}_{n-1}&=&-\displaystyle{\frac{n(n-1)}{2}\frac{f_{t_xt_x}}{f^{n+1}_{t_x}}},
\quad n\geq 2,\\
 b^{(n)}&=&-\displaystyle{\frac{f_t}{f_{t_x}}a^{(n)}_1}, \quad  n\geq 2,
 \end{array}\right.
\end{equation}

\begin{equation}\label{gc10}
a^{(n)}_{n-2}=\frac{(n-2)(n^2-1)n}{4}\frac{f^2_{t_xt_x}}{2f^{n+2}_{t_x}}-
\frac{(n-2)(n-1)n}{3}\frac{f_{t_xt_xt_x}}{2f^{n+1}_{t_x}}, \quad
n\geq 3\, . \end{equation}
\end{lem}

\noindent {\bf{Proof}}. We use induction to prove the Lemma. As
Lemma \ref{gcl1} shows the base of
Mathematical Induction holds.\\
Assuming the representation (\ref{gc8}) for $DK_nD^{-1}$ is true and
all coefficients $a^{(n)}_k$ are functions of $t$, $t_1$, $t_x$
only, consider $DK_{n+1}D^{-1}$. We have,
\begin{eqnarray*}
DK_{n+1}D^{-1}&=&[DXD^{-1}, DK_nD^{-1}]\\
&=&\left[\frac{1}{f_{t_x}}X,
a^{(n)}_nK_n+a^{(n)}_{n-1}K_{n-1}+a^{(n)}_{n-2}K_{n-2}+\ldots+a^{(n)}_1K_1+b^{(n)}X
\right]\\
&=&
a^{(n+1)}_{n+1}K_{n+1}+a^{(n+1)}_{n}K_{n}+a^{(n+1)}_{n-1}K_{n-1}+\ldots+a^{(n+1)}_1K_1+b^{(n+1)}X,
\end{eqnarray*}
where
\begin{eqnarray*}
a^{(n+1)}_{n+1}&=&\frac{1}{f_{t_x}}a^{(n)}_{n}, \\
a^{(n+1)}_{n-k}&=&\frac{1}{f_{t_x}}X(a^{(n)}_{n-k})+\frac{1}{f_{t_x}}a^{(n)}_{n-k-1},
\quad 0\leq k \leq n-2,\\
a^{(n+1)}_1&=&\frac{1}{f_{t_x}}X(a^{(n)}_1).
\end{eqnarray*}
It is easy to see that then $a^{(n+1)}_{n-k}$, $0\leq k\leq n-1$,
are functions of $t$, $t_1$, $t_x$ only as well.

 \noindent Assuming formulas (\ref{gc9}) and (\ref{gc10}) for  $a^{(n)}_n$, $
a^{(n)}_{n-1}$  and $a^{(n)}_{n-2}$ are true, the following
equality
\begin{eqnarray*}
DK_{n+1}D^{-1}&=&a^{(n+1)}_{n+1}K_{n+1}+a^{(n+1)}_{n}K_{n}+a^{(n+1)}_{n-1}K_{n-1}+\ldots+a^{(n+1)}_1K_1+b^{(n+1)}X\\
&=& \left[\frac{1}{f_{t_x}}X   , \frac{1}{f^n_{t_x}}K_n+
a^{(n)}_{n-1}K_{n-1}+a^{(n)}_{n-2}K_{n-2}+\ldots+a^{(n)}_1K_1+b^{(n)}X\right]
\end{eqnarray*}
implies that
\begin{equation*}
 a^{(n+1)}_{n+1}=\frac{1}{f^{n+1}_{t_x}},
\end{equation*}
\begin{eqnarray*}
a^{(n+1)}_{n}&=&\frac{1}{f_{t_x}}X\left(\frac{1}{f^n_{t_x}}\right)+\frac{1}{f_{t_x}}a^{(n)}_{n-1}\\
&=&
-\frac{nf_{t_xt_x}}{f^{n+2}_{t_x}}-\frac{n(n-1)f_{t_xt_x}}{2f^{n+2}_{t_x}}=-\frac{n(n+1)f_{t_xt_x}}{2f^{n+2}_{t_x}},
\end{eqnarray*}
\begin{eqnarray*}
a^{(n+1)}_{n-1}&=&\frac{1}{f_{t_x}}X(a^{(n)}_{n-1})+\frac{1}{f_{t_x}}a^{(n)}_{n-2}\\
&=& \frac{(n-1)n(n+1)(n+2)}{4}\frac{f^2_{t_xt_x}}{2f^{n+3}_{t_x}}-
\frac{(n-1)n(n+1)}{3}\frac{f_{t_xt_xt_x}}{2f^{n+2}_{t_x}}.
\end{eqnarray*}
Using the same notation for $H$ and $H^*$ as in Lemma \ref{gcl1}, by
(\ref{gc4}), we have (for $n\geq 2$),
\begin{eqnarray*}
&&DK_nD^{-1}H=D\left\{X^n(f)\frac{\partial}{\partial t_1
}+X^n(g)\frac{\partial}{\partial
t_{-1}}+X^n(f_1)\frac{\partial}{\partial t_2
}+X^n(g_{-1})\frac{\partial}{\partial t_{-2}
}+\ldots\right\}H^*\\
&&= D\left\{X^n(f)\frac{\partial H^*}{\partial t_{1}
}+X^n(g)\frac{\partial H^*}{\partial t_{-1}}+ X^n(g)
g_{t_{-1}}\frac{\partial H^*}{\partial g}+X^n(f_1)\frac{\partial
H^*}{\partial t_2 }+\ldots\right\}\\
&&= D(X^n(f))\frac{\partial H}{\partial t_2}+
D(X^n(g))\frac{\partial H}{\partial t}+
D(X^n(g))D(g_{t_{-1}})\frac{\partial H}{\partial
t_x}+D(X^n(f_1))\frac{\partial H}{\partial t_3}+\ldots\\&&=
D(X^n(g))\frac{\partial H}{\partial t} -
\frac{f_t}{f_{t_x}}D(X^n(g))X H+ \sum_{k=1}^\infty
\left(\alpha_k^{(n)}\frac{\partial }{\partial
t_k}+\beta_k^{(n)}\frac{\partial }{\partial
t_{-k}}\right)H\\
&&=(b^{(n)}X+a^{(n)}_{1}K_1+a^{(n)}_2K_2+\ldots +a^{(n)}_nK_n)H .
\end{eqnarray*}
Since among $X$, $K_k$, $1\leq k\leq n$ differentiation by    $t$ is
used only in $K_1$ and differentiation by $t_x$ is used only in $X$
then $a^{(n)}_1=D(X^n(g))$ and
$b^{(n)}=-\displaystyle{\frac{f_t}{f_{t_x}}D(X^n(g))}$, that implies
that $b^{(n)}=-\displaystyle{\frac{f_t}{f_{t_x}}a^{(n)}_1}$. The
fact that $b^{(n)}$ is a function of
 $t$, $t_1$ and $t_x$ only follows then from the analogous result for
$a^{(n)}_1$ .
 $\Box$

\begin{lem}\label{new}
Suppose that the vector field
$$K=\sum\limits_{j=1}^\infty\left\{\alpha(k )\frac{\partial }{\partial
t_{k}}+\alpha(-k )\frac{\partial }{\partial t_{-k}}\right\}$$
solves the equation $DKD^{-1}=hK$, where $h$ is a function of
variables $t$, $t_{\pm 1}$, $t_{\pm 2}$, $\ldots$, $t_x$,
$t_{xx}$, $ \ldots$, then $K=0$.
\end{lem}
The proof of Lemma \ref{new} can be easily derived from the
following formula
\begin{eqnarray}\label{DKD-1}
DKD^{-1}&=&-\frac{f_t}{f_{t_x}}D(\alpha(-1))X+D(\alpha(-1))\frac{\partial
}{\partial t}+D(\alpha(-2))\frac{\partial}{\partial t_{-1}}\nonumber \\
&+&\sum\limits_{j=2}^\infty\left\{D(\alpha(j-1))\frac{\partial
}{\partial t_j}+D(\alpha(-j-1))\frac{\partial }{\partial
t_{-j}}\right\}\, .
\end{eqnarray}

 \noindent Consider the linear space $L^*$
generated by $X$ and $K_n$, $n\geq 0$. It is  a subset in the finite
dimensional Lie algebra $L_x$. Therefore, there exists a natural
number $N$ such that
\begin{equation}\label{gc11}
K_{N+1}=\mu X+ \lambda_0 K_0+\lambda_1K_1+\ldots +\lambda_N
K_N\end{equation} and $X$, $K_n$, $0\leq n\leq N$ are linearly
independent.  It can be proved that the coefficients $\mu$,
$\lambda_i$, $0\leq i\leq N$, are functions of finite number of
dynamical variables. Since $\mu=\lambda_0=\lambda_1=0$, then the
equality above should be studied only if $N\geq 2$, or the same, if
the dimension of $L_x$ is 4 or more. Case, when the dimension of
$L_x$ is equal to 3 must be considered separately.

Assume $N\geq 2$. Then
\begin{eqnarray*}
DK_{N+1}D^{-1}&=&D(\lambda_2)DK_2D^{-1}+D(\lambda_3)DK_3D^{-1}+
\ldots +
D(\lambda_{N-1})DK_{N-1}D^{-1}\\&+&D(\lambda_N)DK_ND^{-1}.
\end{eqnarray*}
Rewriting $DK_kD^{-1}$ in the last equation for each $k$, $2\leq
k\leq N+1$, using formulas (\ref{gc8}), and $K_{N+1}$ as a linear
combination (\ref{gc11}) allows us to compare coefficients before
$K_k$, $2\leq k\leq N$ and obtain  the following system of
equations.
\begin{equation}\label{gc12}\begin{array}{l}
a^{(N+1)}_{N+1}\lambda_N+a^{(N+1)}_{N}=D(\lambda_N)a^{(N)}_N\\
\\
a^{(N+1)}_{N+1}\lambda_{N-1}+a^{(N+1)}_{N-1}=D(\lambda_{N-1})a^{(N-1)}_{N-1}+D(\lambda_{N})a^{(N)}_{N-1}
\\
\\
a^{(N+1)}_{N+1}\lambda_{N-2}+a^{(N+1)}_{N-2}=D(\lambda_{N-2})a^{(N-2)}_{N-2}+D(\lambda_{N-1})a^{(N-1)}_{N-2}
+D(\lambda_{N})a^{(N)}_{N-2}\\
\ldots\\
a^{(N+1)}_{N+1}\lambda_{k}+a^{(N+1)}_{k}=D(\lambda_{k})a^{(k)}_{k}+D(\lambda_{k+1})a^{(k+1)}_{k}+\ldots
+D(\lambda_{N})a^{(N)}_{k},
\end{array}
\end{equation}
for $2\leq k\leq N$. Using  the fact that coefficients
$\lambda_k$, $2\leq k\leq N$, depend on a finite number of
arguments, it is easy to see that all of them are functions of
only variables $t$ and $t_x$.

\begin{rem} The first two equations of the system (\ref{gc12}) are
\begin{equation}  \label{gc13}
\begin{array}{l}
\displaystyle{\frac{1}{f^{N+1}_{t_x}}\lambda_N-\frac{N(N+1)}{2}\frac{f_{t_xt_x}}{f^{N+2}_{t_x}}=
\frac{1}{f^{N}_{t_x}}D(\lambda_N)}\, ,\\
\\
\displaystyle{\frac{1}{f^{N+1}_{t_x}}\lambda_{N-1}+\frac{(N^2-1)N(N+2)}{4}\frac{f^2_{t_xt_x}}{2f^{N+3}_{t_x}}-
\frac{(N-1)N(N+1)}{3}\frac{f_{t_xt_xt_x}}{2f^{N+2}_{t_x}}}\\
\\
=\displaystyle{\frac{1}{f^{N-1}_{t_x}}D(\lambda_{N-1})-\frac{N(N-1)}{2}\frac{f_{t_xt_x}}{f^{N+1}_{t_x}}D(\lambda_N)}
\, .
\end{array}
\end{equation}
\end{rem}

\begin{lem}\label{sdl3}
 $K_2=0$ if and only if $f_{t_xt_x}=0$.
 \end{lem}

\noindent {\bf Proof}. Assume  $K_2=0$.   By representation
(\ref{gc4}) we have $X^2(f)=0$, that is  $f_{t_xt_x}=0$.
Conversely, assume that $f_{t_xt_x}=0$. By (\ref{gc6}) we have
$DK_2D^{-1}  = \frac{1}{f^2_{t_x}}K_2$ that implies, by
Lemma~\ref{new}, that $K_2=0$. $\Box$

\noindent Introduce
\begin{equation}\label{sd1}
Z_2=[K_0, K_1].
\end{equation}

\begin{lem} \label{sdl4}
We have,
\begin{equation}\label{sd2}
DZ_2D^{-1}=\frac{1}{f_{t_x}}Z_2-\frac{t_x
f_t+ff_{t_1}}{f^2_{t_x}}K_2+CK_1-\frac{f_t}{f_{t_x}}CX,
\end{equation} where
$$C=-\frac{t_xf_{t_xt}}{f^2_{t_x}}-\frac{ff_{t_xt_1}}{f^2_{t_x}}+
\frac{f_t}{f^2_{t_x}}+\frac{f_{t_1}}{f_{t_x}}+
\frac{t_xf_tf_{t_xt_x}}{f^3_{t_x}}+\frac{ff_{t_1}f_{t_xt_x}}{f^3_{t_x}}\,
.
 $$
\end{lem}

\noindent{\bf Proof}. Using the formulas (\ref{gc5}) and (\ref{gc6})
for $DK_0D^{-1}$, $DK_1D^{-1}$ and the definition (\ref{sd1}) of
$Z_2$ we have,
\begin{eqnarray*}
&&DZ_2D^{-1}=[DK_0D^{-1},DK_1D^{-1}]=[K_0-AX,\frac{1}{f_{t_x}}K_1-BX]\\
&&= K_0\left(\frac{1}{f_{t_x}}\right)K_1+\frac{1}{f_{t_x}}Z_2 -
K_0(B) X+BK_1-AX\left(\frac{1}{f_{t_x}}\right)K_1\\
&&-
A\frac{1}{f_{t_x}}K_2+\frac{1}{f_{t_x}}K_1(A)X+AX(B)X-BX(A)X\\
&&= \frac{1}{f_{t_x}}Z_2 -A\frac{1}{f_{t_x}}K_2+
 \left(K_0\left(\frac{1}{f_{t_x}}\right)+B-
AX\left(\frac{1}{f_{t_x}}\right)\right)K_1\\
&&+ \left( AX(B)-BX(A)-K_0(B)+\frac{1}{f_{t_x}}K_1(A)\right)X ,
\end{eqnarray*}
where
$$
A=\frac{t_xf_t+ff_{t_1}}{f_{t_x}}, \qquad
B=\frac{f_t+f_{t_x}f_{t_1}}{f^2_{t_x}}\,.
$$
The coefficient before $K_1$ is
\begin{eqnarray*}
K_0\left(\frac{1}{f_{t_x}}\right)+B-
AX\left(\frac{1}{f_{t_x}}\right)=-t_x\frac{f_{t_xt}}{f^2_{t_x}}-f\frac{f_{t_xt_1}}{f^2_{t_x}}
+\frac{f_t+f_{t_x}f_{t_1}}{f^2_{t_x}}+\frac{f_{t_xt_x}}{f^2_{t_x}}\frac{t_xf_t+ff_{t_1}}{f_{t_x}}:=C.
\end{eqnarray*}
It follows from (\ref{DKD-1}) that the coefficient before X in
(\ref{sd2}) is $-\displaystyle{\frac{f_t}{f_{t_x}}}C$.
 $\Box$

\bigskip

\begin{lem}\label{sdl5} The dimension of the Lie algebra $L_x$ generated by $X$ and $K_0$ is equal to 3
 if and only if
\begin{equation}\label{sd4}
f_{t_xt_x}=0\end{equation} and
\begin{equation}\label{sd5}
\displaystyle{-\frac{t_xf_{t_xt}}{f^2_{t_x}}-\frac{ff_{t_xt_1}}{f^2_{t_x}}+
\frac{f_t}{f^2_{t_x}}+\frac{f_{t_1}}{f_{t_x}}=0}\, .\end{equation}
\end{lem}

\bigskip

\noindent {\bf Proof}. Assume the dimension of the Lie algebra
$L_x$ generated by $X$ and $K_0$ is equal to 3. It means that the
algebra consists of $X$, $K_0$ and $K_1$ only, and
$K_2=\lambda_1X+\lambda_2K_0+\lambda_3K_1$,
$Z_2=\mu_1X+\mu_2K_0+\mu_3 K_1$ for some functions $\lambda_i$ and
$\mu_i$. Since among $X$, $K_0$, $K_1$, $K_2$ and $Z_2$ we have
differentiation by $t_x$ only in $X$, differentiation by $x$ only
in $K_0$, then $\lambda_1=\lambda_2=\mu_1=\mu_2=0$. Therefore,
$K_2=\lambda_3K_1$ and $Z_2=\mu_3K_1$.  Also, among $K_1$, $K_2$
and $Z_2$ we have differentiation by $t$ only in $K_1$ then
$\lambda_3=\mu_3=0$. We have proved that if the dimension of the
Lie algebra $L$ is 3 then $K_2=0$ and $Z_2=0$. By
Lemma~\ref{sdl3}, condition (\ref{sd4}) is satisfied. It follows
from (\ref{sd2}) that
$$
0=DZ_2D^{-1}= \frac{1}{f_{t_x}}Z_2-\frac{t_x
f_t+ff_{t_1}}{f^2_{t_x}}K_2+CK_1-\frac{f_t}{f_{t_x}}CX=CK_1-\frac{f_t}{f_{t_x}}CX.
$$
Since $X$ and $K_1$ are linearly independent then  equality
$CK_1-\displaystyle{\frac{f_t}{f_{t_x}}}CX=0$ implies $C=0$. Equality (\ref{sd5}) follows from (\ref{sd4}) and $C=0$. \\
Conversely, assume that properties (\ref{sd4}) and (\ref{sd5}) are
satisfied. To prove that the dimension of the Lie algebra $L_x$ is
equal to 3 it is enough to show that $K_2=0$ and $Z_2=0$. It
follows from (\ref{sd4}) and Lemma \ref{sdl3}  that  $K_2=0$. From
the formula (\ref{sd2}) for $DZ_2D^{-1}$,  property (\ref{sd5})
and knowing that $K_2=0$ we have that
$DZ_2D^{-1}=\displaystyle{\frac{1}{f_{t_x}}}Z_2$ that implies, by
Lemma~\ref{new}, that $Z_2=0$. $\Box$

\section{Equations with characteristic algebras of the minimal possible dimensions.}

\begin{cor}
If Lie algebras for $n-$ and $x-$ integrals have
 dimensions 2 and 3 respectively, then equation
 $t_{1x}=f(t,t_1, t_x)$ can be reduced to $t_{1x}=t_x+t_1-t$.
\end{cor}
\noindent {\bf Proof}. By Lemma \ref{sdl5} and Corollary
\ref{cornint}, the dimensions of $n$- and $x$-Lie algebras are 2 and
3 correspondingly means equations (\ref{Y2=0relation}), (\ref{sd4}),
and (\ref{sd5}) are satisfied. It follows from property (\ref{sd4})
that $f(t,t_1,t_x)=G(t,t_1) t_x+H(t,t_1) $ for some functions
$G(t,t_1)$ and $H(t,t_1)$. By (\ref{Y2=0relation}),
$G_tt_x+H_t+\{D^{-1}(G_{t_1}t_x+H_{t_1})\}G=0$, that is
\begin{equation}
\label{sd6}D^{-1}(G_{t_1}t_x+H_{t_1})=-\frac{G_t}{G}t_x-\frac{H_t}{G}
\, .\end{equation} Note that $t_{1x}=Gt_x+H$ implies
$t_x=D^{-1}(G)t_{-1x}+D^{-1}(H)$ and, therefore,
$t_{-1x}=\frac{1}{D^{-1}(G)}t_x-\frac{D^{-1}(H)}{D^{-1}(G)}$. We
continue with (\ref{sd6}) and obtain the following equality
$$
D^{-1}\left(\frac{G_{t_1}}{G}\right)t_x-D^{-1}\left(\frac{G_{t_1}H}{G}\right)+D^{-1}(H_{t_1})=
-\frac{G_t}{G}t_x-\frac{H_t}{G}
$$
that gives rise to two equations
\begin{equation}\label{sd7}
D^{-1}\left(\frac{G_{t_1}}{G}\right)=-\frac{G_t}{G}, \qquad
D^{-1}\left(H_{t_1}-\frac{G_{t_1}H}{G}\right)=-\frac{H_t}{G}  \, .
\end{equation}
It is seen from the first equation of (\ref{sd7}) that
$\displaystyle{\frac{G_t}{G}}$ is a function that depends only on
variable $t$, even though functions $G$ and $G_{t}$ depend on
variables $t$ and $t_1$. Denote
$\displaystyle{\frac{G_t}{G}}=:a(t)$. Then
$\displaystyle{\frac{G_{t_1}}{G}}=-a(t_1)$. The last two equations
imply that $G=A_1(t_1)e^{\tilde{a}(t)}=A_2(t)e^{-\tilde{a}(t_1)}$
for some functions $A_1(t_1)$ and $A_2(t)$ and
$\tilde{a}(t)=\int_0^ta(\tau)d\tau$. Noticing that
$A_1(t_1)e^{\tilde{a}(t_1)}=A_2(t)e^{-\tilde{a}(t)}$ we conclude
that $A_1(t_1)e^{\tilde{a}(t_1)}$ is a constant.  Denoting
$\gamma:=A_1(t_1)e^{\tilde{a}(t_1)}$ and
$G_1(t):=e^{-\tilde{a}(t)}$
 we have
 \begin{equation}\label{sd8}
 G(t,t_1) =\gamma\frac{G_1(t_1)}{G_1(t)}\qquad {\mbox{and, therefore}},
 \qquad f(t,t_1,t_x)=\gamma\frac{G_1(t_1)}{G_1(t)}t_x+H.
 \end{equation}
The second equation of (\ref{sd7}) implies that
\begin{equation}\label{sd9}
\frac{H_t}{G}=-\mu(t) \quad {\mbox{and}} \quad
H_{t_1}-\frac{G_{t_1}H}{G}=\mu(t_1)
\end{equation}
 for some function $\mu(t)$.
Using (\ref{sd8}), the second equation in (\ref{sd9}) can be
rewritten as
$H_{t_1}-\displaystyle{\frac{G'_1(t_1)H}{G_1(t_1)}}=\mu(t_1)$, or
the same, as
$\left\{\displaystyle{\frac{H(t,t_1)}{G_1(t_1)}}\right\}_{t_1}=\displaystyle{\frac{\mu(t_1)}{G_1(t_1)}}$.
It means that
\begin{equation}\label{sd10}
H(t,t_1)=G_1(t_1)H_1(t_1)+G_1(t_1)H_2(t)
\end{equation}
for some functions $H_1(t_1)$ and $H_2(t)$. By substituting
$H(t,t_1)$ from (\ref{sd10}) and $G(t,t_1)$ from (\ref{sd8})  into
the second equation of (\ref{sd9}) we have,
\begin{eqnarray*}
&&G'_1(t_1)H_1(t_1)+G_1(t_1)H'_1(t_1)+G'_1(t_1)H_2(t)\\
&&\hspace{36mm}-
\frac{G'_1(t_1)}{G_1(t_1)}(G_1(t_1)H_1(t_1)+G_1(t_1)H_2(t))=\mu(t_1).
\end{eqnarray*}
After all cancellations it becomes
\begin{equation}
\label{sd11} G_1(t_1)H'_1(t_1)=\mu(t_1), \quad {\mbox{or the
same}}, \quad G_1(t)H'_1(t)=\mu(t)\, .\end{equation} By
substituting $G(t,t_1)$ from (\ref{sd8}) and $H(t,t_1)$ from
(\ref{sd10}) into the first equation of (\ref{sd9}) we have,
\begin{equation}\label{sd12}
H'_2(t)G_1(t)=-\gamma\mu(t)  \, .
\end{equation}
Combining together (\ref{sd11}) and (\ref{sd12}) we obtain that
$H'_2(t)G_1(t)=-\gamma G_1(t)H'_1(t)$, or the same,
$H'_2(t)=-\gamma H'_1(t)$, or $(H_2(t)+\gamma H_1(t))'=0$ that
implies that $H_2(t)=-\gamma H_1(t)+\eta$ for some constant
$\eta$. Therefore,
\begin{equation}\label{sd13}
f(t,t_1,
t_x)=\gamma\frac{G_1(t_1)}{G_1(t)}t_x+G_1(t_1)H_1(t_1)-\gamma
G_1(t_1)H_1(t)+\eta G_1(t_1)  \, .
\end{equation}
Note that only properties (\ref{sd4}) and (\ref{Y2=0relation})
were used to obtain representation (\ref{sd13}) for
$f(t,t_1,t_x)$. Using (\ref{sd5}) and (\ref{Y2=0relation}) we have
\begin{eqnarray*}
&&0=t_x\left\{\gamma\frac{G_1(t_1)}{G_1^2(t)}G'_1(t)\right\}-
\gamma \frac{G_1(t_1)}{G^2_1(t)}G'_1(t)t_x-\gamma G_1(t_1)H'_1(t)
\\&&-\gamma\frac{G'_1(t_1)}{G_1(t)} \left\{\gamma
\frac{G_1(t_1)}{G_1(t)}t_x+G_1(t_1)H_1(t_1)-\gamma
G_1(t_1)H_1(t)+\eta G_1(t_1)\right\}\\
&&+\gamma\frac{G_1(t_1)}{G_1(t)}\left\{\gamma
\frac{G'_1(t_1)}{G_1(t)}t_x+
G'_1(t_1)H_1(t_1)+G_1(t_1)H'_1(t_1)-\gamma G'_1(t_1)H_1(t) +\eta
G'_1(t_1)\right\}\\
&&= \frac{\gamma
G_1(t_1)}{G_1(t)}\left\{-H'_1(t)G_1(t)+G_1(t_1)H'_1(t_1)\right\},
\end{eqnarray*}
i.e. $-H'_1(t)G_1(t)+G_1(t_1)H'_1(t_1)=0$. This implies that
$H'_1(t)G_1(t)=c$, where $c$ is some constant. Substituting
$G_1(t)=\displaystyle{\frac{c}{H'_1(t)}} $ into (\ref{sd13}) we
have,
\begin{equation}
\label{sd14} f(t,t_1,t_x)=\gamma
\frac{H'_1(t)}{H'_1(t_1)}t_x+c\frac{H_1(t_1)}{H'_1(t_1)}-\gamma c
\frac{H_1(t)}{H'(t_1)}+\eta \frac{c}{H'_1(t_1)}   \, .
\end{equation}
By using substitution $s=H_1(t)$ equation (\ref{sd14}) is reduced
to $ s_{1x}=\gamma s_x +cs_1-c\gamma s+\eta c$. Introducing
$\tilde{x}=cx$ allows to rewrite the last equation as
\begin{equation}\label{sd15}
s_{1\tilde{x}}=\gamma s_{\tilde{x}} +s_1-\gamma s +\eta \, .
\end{equation}
 If $\gamma =1$ substitution $s=\tau-n\eta$ reduces (\ref{sd15}) to
 $\tau_{1\tilde{x}}=\tau_{\tilde{x}}+\tau_1-\tau$. If $\gamma\ne 1$, substitution
 $s=\gamma^n\tau+\eta\displaystyle{\frac{\gamma^{n}-1}{1-\gamma}}$ reduces (\ref{sd15}) to
 $\tau_{1\tilde{x}}=\tau_{\tilde{x}}+\tau_1-\tau$.
 $\Box$

\section*{Acknowledgments}
The authors thank Prof. M. G\"{u}rses for fruitful discussions.
Two of the authors (AP, NZ) thank the Scientific and Technological
Research Council of Turkey (TUB{\.{I}}TAK) and the other (IH)
thanks (TUB{\.{I}}TAK), the Integrated PhD. Program (BDP) and
grants RFBR $\#$ 05-01-00775 and RFBR $\#$ 06-01-92051-CE$\_$a for
partial financial support.

\end{document}